\documentclass[aps,prb, twocolumn,showpacs,groupedaddress]{revtex4}
\usepackage{graphicx}
\usepackage{amsmath,amsfonts,amssymb}
\usepackage{bm}
\usepackage{subfigure}

\begin{document}

\title{Structural instabilities and sequence of phase transitions in  SrBi$_2$Nb$_2$O$_9$ and SrBi$_2$Ta$_2$O$_9$ from first principles and Monte Carlo simulations}

\author{Urko Petralanda}
\author{I. Etxebarria}
\email[]{inigo.etxebarria@ehu.es}
\affiliation{Fisika Aplikatua II Saila, Zientzia eta Teknologia Fakultatea, Euskal Herriko Unibertsitatea, P.K. 644, 48080 Bilbao, Spain.}

\renewcommand{\vec}[1]{\mathbf{#1}}
\begin{abstract}

Despite their structural similarities,  SrBi$_2$Ta$_2$O$_9$ (SBT) and SrBi$_2$Nb$_2$O$_9$ (SBN)  undergo a different sequence of phase transitions.  The phase diagram of  SBT as a function  of the temperature includes an intermediate phase between the high-temperature phase and the ferroelectric ground state, while in the Niobium compound the intermediate phase is suppressed and a single  transition between the high- and low temperature structures is observed.  We present $\emph{ab initio}$ calculations that  reveal the relevance of a trilinear coupling between  three symmetry-adapted modes to stabilize the ground sate in both compounds,   being this coupling much stronger in SBN. Within the framework of the phenomenological Landau theory, it is shown that by solely increasing the strength of the trilinear coupling  the topology of the phase diagram of SBT can change up to suppress the intermediate phase. Monte Carlo simulations on an idealized $\phi^4$ Hamiltonian confirm that the trilinear coupling is the key parameter that determines the sequence of phase transitions and  that for higher dimensionality of the order parameters  the   stability region of the intermediate phase is narrower.
\end{abstract}

\pacs{77.80.-e, 77.84.-s, 63.20.Ry, 71.15.Mb}

\maketitle

\section{\label{sec:intro} Introduction}
\begin{figure}
\includegraphics[scale = 0.50]{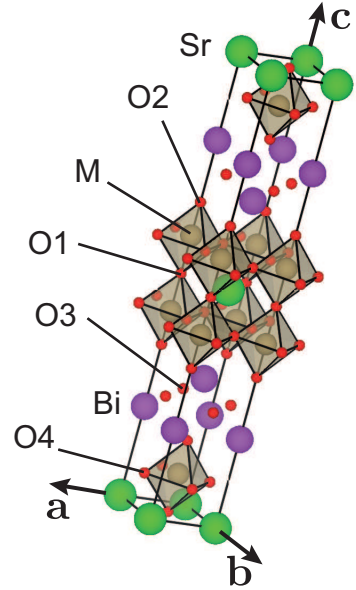} 
		\caption{Structure of SBT and SBN in the tetragonal phase (M=Ta,Nb).}
\label{fig:structure}
\end{figure}

The Aurivillius phases are layered bismuth compounds that  obey the general formula Bi$_{2m}$A$_{n-m}$B$_n$O$_{3_{n+m}}$ \cite{Aurivillius}. The family includes many members that present ferroelectricity at room temperature, and  they have been widely studied for  potential technological applications mainly in thin film nonvolatile memories.\cite{Appli1, Appli2,Appli3} SrBi$_2$Ta$_2$O$_9$ (SBT) and  SrBi$_2$Nb$_2$O$_9$ (SBN)  are members of the family, with $n=2$ and $m=1$ and their structure is formed by alternating two SrMO$_3$ (M$=$Ta, Nb) perovskite blocks and one Bi$_2$O$_2$ slab (Fig. \ref{fig:structure}).

Tantalum and Niobium  present very similar physical and chemical properties, including the valence and  atomic radii. As a result, SBT and SBN  present isomorphous polar  structures at room temperature \cite{Rae,Shimakawa,SBT_transitions,Ismunandar,Sneeden,Boullay} and analogous mechanical and electrical  characteristics. However, their phase diagrams are qualitatively different: on increasing the temperature SBT undergoes a phase transition to a non-polar orthorhombic phase that does not arise in SBN. \cite{Sneeden,Boullay,SBT-PT}

Previous first-principles calculations by Perez-Mato \emph{et al} revealed that the trilinear coupling of two primary unstable modes and a secondary hard mode is critical to stabilize the ground state of SBT. \cite{Manu} This work provided a simple and plausible scheme  in contrast with others alternatives  where the eventual existance of negative biquadratic coupling triggers the simultaneous condensation of several modes. \cite{Holakowsky} Later, the key role of the trilinear coupling has been found in several other compounds\cite{Ghosez_nature,Tril1,Tril2,Tril3,Tril4,PZO-PRL,PZO-Iniguez} and its influence has been proposed as the origen of the so-called \emph{avalanche}\cite{pmato2008}  phase transitions and the \emph{hybrid} improper ferroelectricity.\cite{Tril1}

In this work we revisit the SBT and extend the \emph{ab initio} study to SBN in order to investigate the role of the trilinear coupling in the stabilization of the ferroelectric structure and its influence in the sequences of phase diagrams.

\section{\label{sec:bi} Symmetry considerations, distortions and phase transitions }
The high- and low-temperatures structures of SBT and SBN are isomorphous:  tetragonal with space group $I4/mmm$ (No. 139) at high enough temperatures\cite{Rae,Sneeden} and orthorhombic $A2_1am$ (No. 36) at low temperatures. In the case of SBN there is a direct phase transition from the tetragonal to the ferroelectric structure at 800K,\cite{Boullay, Sneeden} whereas SBT presents an intermediate nonpolar orthorhombic $Amam$ phase (No. 63) between 500-600K and 770K. \cite{Rae}. The cell parameters of the two orthorhombic phases  are related to the tetragonal ones according to $\vec{c_O}=\vec{c_T}$, $\vec{a_O}=\vec{a_T}+\vec{b_T}$ and $\vec{b_O}=\vec{a_T}-\vec{b_T}$, which  means that the volume of the unit cell is doubled in the tetragonal-to-orthorhombic  transitions.

The tetragonal space group ($I4/mmm$) is a supergroup of the two orthorhombic space groups, the intermediate phase of SBT ($Amam$) and the ground state of both compounds ($A2_1am$). In consequence, the low temperature structures can be described in terms of symmetry-adapted distortions with respect to the parent tetragonal structure. \cite{symmetry} Table \ref{tab:mode_decomposition} shows this decomposition in terms of the symmetry-adapted modes that break the tetragonal symmetry  for the  experimental structures of SBT and SBN  obtained with the aid of the tool AMPLIMODES  (www.cryst.ehu.es). \cite{amplimodes} The hierarchy of the modes is similar in both compounds:  the amplitude of $X _{3}^{-}$ is dominant and can be considered as the primary order parameter and the $X _{2}^{+}$ distortion seems to play a secondary role. The analysis of the $\Gamma _{5}^{-}$ polar distortion is more delicate because due to the polar character of the ground state its amplitude depends on the origin. In Table \ref{tab:mode_decomposition} the chosen origin does not change the arithmetic center of the parent structure.
According to the isotropy subgroups the intermediate phase of SBT is naturally explained as a condensation of the $X _{3}^{-}$ mode, and in terms of the Landau theory of phase transitions the $I4/mmm \rightarrow Amam$ transition can be continuous. The second transition to the ferroelectric structure should then consist on the  condensation of the polar mode. In the case of SBN a direct transition between the tetragonal and the ferroelectric phase  can be  solely described  by a simultaneous condensation of at least any pair of the three modes,  and according to Landau theory, it should be first order: the three relevant modes condense simultaneously in an \emph{avalanche} phase transition \cite{pmato2008}.

The orthorhombic strain is usually very small in the Aurivillius compounds and SBT is not an exception. The experimental  values of cell parameters  in SBT are $a_T=b_T= 3.917$~\AA\; and $c_T=25.114$~\AA\;    for the tetragonal structure, \cite{Sneeden} and  $a_O=5.531$ \AA\;,  $b_O=5.534$ \AA\; and $c_O=24.984$~\AA\; for the ferroelectric phase.\cite{Rae} The deformation of the cell is very small with  $\sqrt{2}a_T/c_T \sim a_O/c_O \sim b_O/c_O$  (all within a 0.5 $\%$ margin), and  the influence of strain can be considered as a residual effect. Up to our knowledge, there is not  such an experimental data available for the tetragonal cell of SBN, and the orthorhombic cell parameters are $a_O=5.519$ \AA\; and $b_O=5.515$ \AA\; and $c_O=25.112$~\AA. \cite{Ismunandar}

A complete description of the atomic displacements associated with the three modes can be found in Ref. \onlinecite{Manu}. In summary, among the 18 degrees of freedom that  constitute a generic distortion that links the tetragonal and ferroelectric structure, seven correspond to  $X_3^-$, and they mainly account for tiltings of the oxygen octahedra  around the $(1,1,0)_T$ direction of the tetragonal cell and displacements of the Bi cations along the $(1,-1,0)_T$ direction. The octahedra behave as  rigid units and, in consequence, neighboring octahedra show antiphase tiltings. The polar $\Gamma_5^-$ distortion involves essentially an antiphase displacement of the Bi atoms and the perovskite blocks, along the  $(1,1,0)_T$ direction. Among the eight independent displacements associated  with the  $X_2^+$ distortion the main amplitudes correspond to  tiltings of the oxygen octahedra around the  $z$ axis and antiphase displacements  of the oxygens  in the  Bi$_2$O$_2$ slabs along the $(1,1,0)_T$ direction.


\begin{table}
\begin{ruledtabular}

\begin{tabular}{cccccccc}
 & & & \multicolumn{5}{c}{Amplitude (\AA)}   \\ \cline{4-8}
 & & Isotropy &\multicolumn{3}{c}{SBT} &\multicolumn{2}{c}{SBN}     \\   \cline{4-6} \cline{7-8}
Irrep&Direction& subgroup & exp.\cite{Rae} & exp.\cite{Shimakawa} & calc. & exp. & calc.\\ \colrule
$X _{3}^{-}$& $(a,-a)$ &$Amam(63)$&0.90&0.86 &1.12&0.81 & 1.14\\
 & &  &(0.99)& (0.99)& & (0.99)  &    \\   
$\Gamma _{5}^{-}$& $(a,a)$& $F2mm(46)$&0.51&0.63&1.01&0.74 & 1.03\\
 & &  &(0.94)& (0.98)& & (0.81)  &    \\   
$X _{2}^{+}$&$(a,-a)$&$Abam(64)$&0.26&0.39&0.56&0.63 & 0.75\\
 & &  &(0.84)& (0.99)& & (0.99)  &    \\   
\end{tabular}
\caption{Summary of the mode decomposition of the experimental and  calculated  distortions of the ferroelectric phases of SBT and SBN  with respect to their corresponding parent (calculated and experimental) structures. Amplitudes are given for modes normalized within the primitive unit cell of the polar structure. The second row shows in parentheses the scalar product of the calculated and experimental polarization vectors after normalization.}
\label{tab:mode_decomposition}
\end{ruledtabular}
\end{table}

\section{\label{sec:hiru} COMPUTATIONAL DETAILS}
The WIEN2k code \cite{wien2k}, based on the full potential LAPW+lo method, was employed for the \emph{ab initio} calculations. Exchange and correlation effects were treated within the GGA approximation with the Perdew-Burke-Ernzehof parameterization \cite{perdew}. The $RK_{max}$ parameter, which is related to the number of radial basis functions used to describe the the inside spheres, was chosen to be 7.5 for both compounds. Calculations in the tetragonal symmetry were performed using a Monkhorst-Pack $k$ point mesh of $8\times8\times8$, which is equivalent to 56 independent $k$ points in the irreducible Brillouin wedge. For calculations in the orthorhombic symmetry a $k$-mesh of $5\times5\times5$, representing 27 independent $k$ points in the IBZ, was chosen in order to keep the $k$ point density as constant as possible. In the case of SBT the radii of the atomic spheres chosen for the calculations were 2.0(Sr), 2.26(Bi), 1.8(Ta), 1.6(O) bohrs in the tetragonal structure and 2.25(Sr), 2.3(Bi), 1.8(Ta), 1.6(O) bohrs in the orthorhombic $A2_1am$ phase. For SBN, the radii were 2.26(Sr), 2.26(Bi), 1.8(Nb), 1.66(O) bohrs for both the tetragonal and the orthorhombic basis. The choice of the parameters was preceded by energy difference convergence tests which confirmed their validity. The convergence criterion for the SCF calculations was of 0.0001 Ry for energy and 0.1 mRy/bohr for forces. 

First, relaxations of the parent structures and  ground states  for the two compounds were carried out. In the case of SBT forces remained below 0.02 mRy/bohr and  0.15mRy/bohr in the tetragonal  and orthorhombic phases respectively. For SBN, the tetragonal relaxation accepted an accuracy under 0.1 mRy/bohr and all forces present in the orthorhombic relaxed structure were under 0.17mRy/bohr. 

As mentioned in the previous section the orthorhombic strain is very small in the Aurivillius compounds. In consequence its effect was neglected and all the calculations were performed  assuming a fixed  cell with tetragonal metric.  The  $c$ parameter of the idealized cell was fixed to the experimental value of the ferroelectric phase and the $a$ and $b$ lattice constants were forced to be equal and preserve the experimental volume of the same phase.

\section{\label{sec:lau} Energetics and couplings}

First, the atomic positions of both compounds were relaxed under tetragonal and  orthorhombic $A2_1am$ symmetries, and the distortions of the low-temperature structure were analyzed in terms of   symmetry modes. As shown in Table \ref{tab:mode_decomposition}, the calculated amplitudes are systematically larger than the experimental values, what could  be related to some degree of disorder in the experimental structures that is not present in the calculations. Table  \ref{tab:mode_decomposition} also shows  the the scalar product of the calculated and experimental polarization vectors after normalization. The extremely good agreement indicates that, apart from a global amplitude,  the calculated and experimental relative atomic displacements are essentially the same.


\begin{table}
\begin{ruledtabular}
\begin{tabular}{ccccc}
\hline
Compound&Abs. minimum&$X _{3}^{-}$&$\Gamma _{5}^{-}$&$X _{2}^{+}$\\
SBT&15.62& 11.72&5.24&1.52\\
SBN&18.34&10.40&3.68&3.52\\
\end{tabular}
\label{tab:wells}
\caption{Depths of the energy wells per formula unit (in mRy) of the ferroelectric ground state (first column)  and  the  pure distortions $X _{3}^{-}$, $\Gamma _{5}^{-}$ and $X _{2}^{+}$ with respect to the relaxed tetragonal parent configuration. }
\end{ruledtabular}
\end{table}

\begin{figure*}
\includegraphics[scale = 0.50]{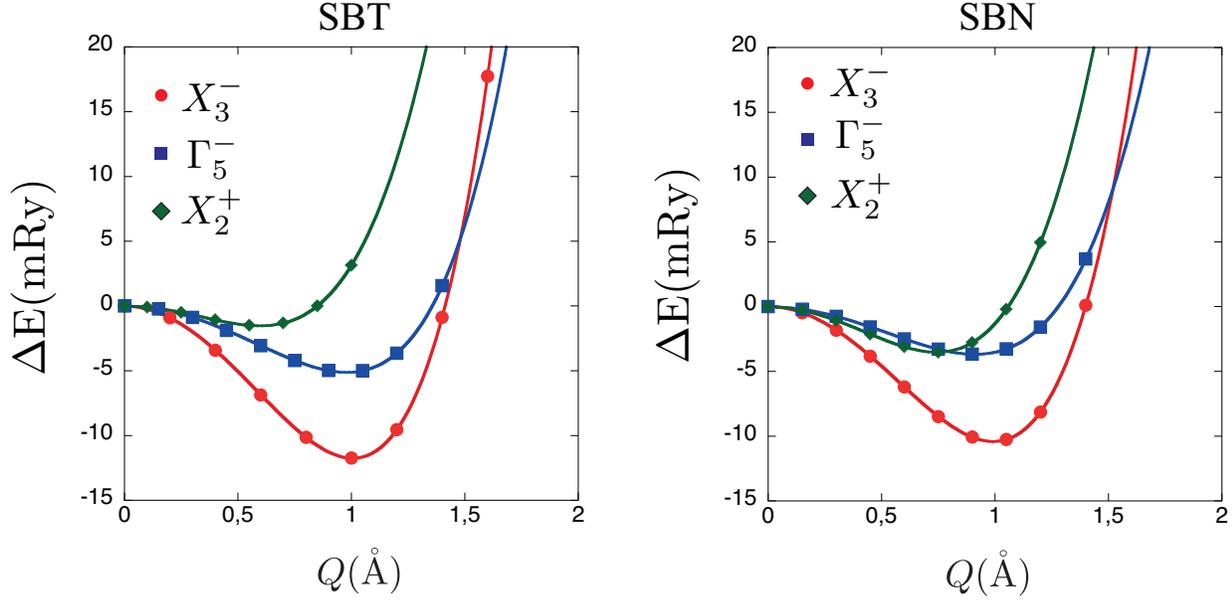} 
		\caption{Energy per formula unit relative to the tetragonal structure of SBT (left) and SBN (right) in terms of the amplitudes  of the $X _{3}^{-}$, $\Gamma _{5}^{-}$ and $X _{2}^{+}$ distortions.}
\label{fig:puremodes}
\end{figure*}

The energy gains of the polar phase with respect to the tetragonal one, together with the depth of the energy wells associated with the pure distortions are listed in Table \ref{tab:wells}. The primary character of the $X_3^-$ order parameter is manifest in both compounds. However, the  role of the other two distortions seems to be different in both compounds. In SBT, $\Gamma _{5}^{-}$ presents a well defined energy minimun, while  the $X _{2}^{+}$ mode is slightly soft  at low amplitudes and  rapidly hardens for medium amplitudes (Fig. \ref{fig:puremodes}) reinforcing its secondary character. In SBN, the energy wells for the  $\Gamma_{5}^{-}$ and $X _{2}^{+}$ modes are very similar  and a clear hierarchy between both distortions cannot be established (Fig. \ref{fig:puremodes}).

The energy variations around the tetragonal configuration in terms of the amplitudes of the three relevant modes can be described by a  polynomial composed of symmetry invariant terms. The knowledge of the coefficients in the energy expansion allows the quantification of the energy for a general distortion and the strength of the couplings between modes. The energetic contribution of a general combination of  $X_3^-$ ,  $\Gamma _{5}^{-}$ and  $X _{2}^{+}$ can be expressed up to fourth order by:
\begin{eqnarray}
\Delta E&=&E_{X_3^-}+E_{\Gamma_5^-}+E_{X_2^+}+E_{X_3^-\Gamma_5^-} \nonumber \\
&&+E_{X_3^-X_2^+}+E_{\Gamma_5^-X_2^+}+E_{X_3^-\Gamma_5^-X_2^+}
\end{eqnarray}
The energy due to the pure modes is given by:
\begin{eqnarray*}
E_{X_3^-}=\frac{1}{2}\kappa_{X_3^-}Q_{X_3^-}^2+\beta_{X_3^-}Q_{X_3^-}^4\\E_{\Gamma_5^-}=\frac{1}{2}\kappa_{\Gamma_5^-}Q_{\Gamma_5^-}^2+\beta_{\Gamma_5^-}Q_{\Gamma_5^-}^4\\E_{X_2^+}=\frac{1}{2}\kappa_{X_2^+}Q_{X_2^+}^2+\beta_{X_2^+}Q_{X_2^+}^4
\end{eqnarray*}
The biquadratic couplings read:
\begin{eqnarray*}
E_{X_3^-\Gamma_5^-}=\delta_{X_3^-\Gamma_5^-}Q_{X_3^-}^2Q_{\Gamma_5^-}^2\\E_{X_3^-X_2^+}=\delta_{X_3^-X_2^+}Q_{X_3^-}^2Q_{X_2^+}^2\\E_{\Gamma_5^-X_2^+}=\delta_{\Gamma_5^-X_2^+}Q_{\Gamma_5^-}^2Q_{X_2^+}^2
\end{eqnarray*}
and the trilinear coupling is:
\begin{eqnarray*}
E_{X_3^-\Gamma_5^-X_2^+}=\gamma_{X_3^-\Gamma_5^-X_2^+}Q_{X_3^-}Q_{\Gamma_5^-}Q_{X_2^+}
\end{eqnarray*}

The coefficients of the polynomials for both compounds were determined by fitting the \emph{ab initio} energies calculated in more than 60 points of the configuration space for each compound (Table \ref{tab:stiff}). The most remarkable difference between both compounds is the magnitude of the trilinear coupling, its value in SBN being much larger than in SBT. The magnitudes of the rest of coefficients are quite similar although the stiffness constants of $\Gamma_5^-$ and X$_2^+$ modes approximately exchange their values and, in consequence, their role as the least unstable distortion.

\begin{figure*}
\includegraphics[scale = 0.45]{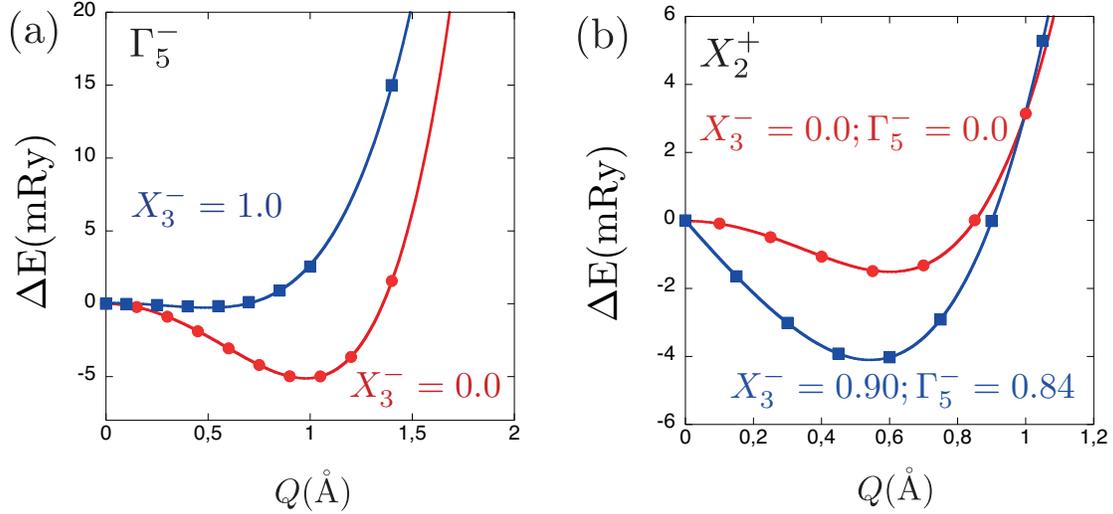} 
	\caption{ Energy per formula unit of SBT along selected lines of the configuration space: (a) in terms of the amplitudes of the the pure polar mode (squares),  and the same mode after the  $X_3^-$ distortion has been frozen to 1\;\AA \; (circles). b) in terms of the amplitude of the  pure $X_2^+$ mode (squares) and the same mode after the amplitudes of  $X_3^-$ and $\Gamma_5^-$ distortions have been fixed to 0.90\;\AA \; and 0.84\;\AA\; respectively (circles). The plots have been shifted vertically to fix a common origin.}
\label{fig:trigger}
\end{figure*}

The condensation of two of the three relevant symmetry modes is enough to lower the tetragonal symmetry down to the ferroelectric space group. However, in both compounds the simultaneous freezing of any pair of modes is penalized by the strong and positive biquadratic couplings $\delta$. Fig \ref{fig:trigger}(a) shows the renormalization of the energies due to the biquadratic coupling for SBT:  the presence of a nonzero $\Gamma _{5}^{-}$ distortion stabilizes the soft $X_3^-$ mode and a simultaneous condensation of both modes becomes  energetically unfavourable. The inclusion of the $X_2^+$ mode is essential to explain the polar ground state of SBT. In fig \ref{fig:trigger}(b) is shown the energy of SBT as a function of the amplitude of the pure  X$_2^+$ mode  and when a mixed distortion with  $Q_{X_3^-}$=0.90\;\AA\ and $Q_{\Gamma_5^-}$=0.84\;\AA\; is applied; the slightly hard X$_2^+$ mode becomes strongly unstable  when the two primary modes  condense. This is a manifestation of the critical relevance of the trilinear coupling between the three modes in the stabilization of the ground state. The  analysis of the couplings in the case of SBN yields the same qualitative picture. 

Analogous conclusions were obtained in Ref. \onlinecite{Manu} for the case of SBT, i.e., the trilinear coupling between  three modes is determinant in the stabilization of the ferroelectric structure.   However, the authors studied the couplings between  modes that correspond to the eigenvectors of the force constant matrix. Particularly, the displacement pattern associated with the $X_2^+$ static distortion of this work involves contributions from two phonons, one hard and one soft. In consequence,  the values of  the coupling constants in Table \ref{tab:stiff} and some details of Figures \ref{fig:puremodes} and \ref{fig:trigger} cannot be directly compared with the results of Ref. \onlinecite{Manu}. A similar scenario has been found in several other compounds, \cite{Tril1, Tril2, Tril3, Tril4} suggesting that the critical role of  trilinear couplings  to stabilize the ground state could be rather common.

\begin{table}
\begin{ruledtabular}
\begin{tabular}{ccc}
 $$&SBT&SBN\\
 \hline
$\kappa_{X _{3}^{-}}(\text{mRy/bohr}^{2})$ & -44.26 & -40.42 \\
$\kappa_{\Gamma _{5}^{-}}(\text{mRy/bohr}^{2})$& -21.41 &-17.28 \\
$\kappa_{X _{2}^{+}}(\text{mRy/bohr}^{2})$& -16.39 & -23.49 \\
$\beta_{X _{3}^{-}}(\text{mRy/bohr}^{4})$ & 11.03 & 10.1    \\
$\beta_{\Gamma _{5}^{-}}(\text{mRy/bohr}^{4}) $& 5.62 & 5.40\\
$\beta_{X _{2}^{+}}(\text{mRy/bohr}^{4}) $& 11.50 & 10.25  \\
${\delta }_{X _{3}^{-}\Gamma _{5}^{-}} (\text{mRy/bohr}^{4})$ & 6.20& 7.76\\
${\delta }_{X _{3}^{-}X _{2}^{+}}(\text{mRy/bohr}^{4})$& 10.60 & 12.60  \\
${\delta }_{X _{2}^{+}\Gamma _{5}^{-}}(\text{mRy/bohr}^{4}) $ & -2.30& 3.83  \\
$\gamma_{{Q}_{X _{3}^{-}}{Q}_{\Gamma _{5}^{-}}{Q}_{X _{2}^{+}}}(\text{mRy/bohr}^{3})$ &-13.81 &-22.55  \\
\end{tabular}
\caption{Polynomial coefficients of the energy expansion of SBT and SBN  obtained by least-square fits. }
\label{tab:stiff}
\end{ruledtabular}

\end{table}

\section{\label{sec:bost} Approximate phase diagram}

The expansion of the energy  in the previous section can be considered  as the zero temperature free energy of the system, and a phenomenological free energy can be approximated under the assumptions of the Landau theory of phase transitions. Then, the temperature renormalization is solely contained  in the quadratic terms and the stiffness constants present a linear dependence on temperature, such that $\kappa_i=a_i(T-T_{0,i})$ for a given $i$ mode, where $a$ and $T_{0,i}$ are constants. Thus, the  \emph{ab initio}  stiffness constants  are related to the transition temperatures and renormalization constants of the modes by $\kappa_{i}=-a_i T_{0,i} $. In a first  approximation the temperature dependence of  higher order terms can be neglected  and the free energy reads:

\begin{eqnarray*}
\begin{split}
F=\sum_i^3\left(\frac{a_i}{2}(T-T_{0.i})Q_i^2+\beta_iQ_i^4\right)\\+E_{X_3^-\Gamma_5^-}+E_{X_3^-X_2^+}+E_{\Gamma_5^-X_2^+}+E_{X_3^-\Gamma_5^-X_2^+}
\end{split}
\end{eqnarray*}

\begin{figure*}
\includegraphics[scale=0.36]{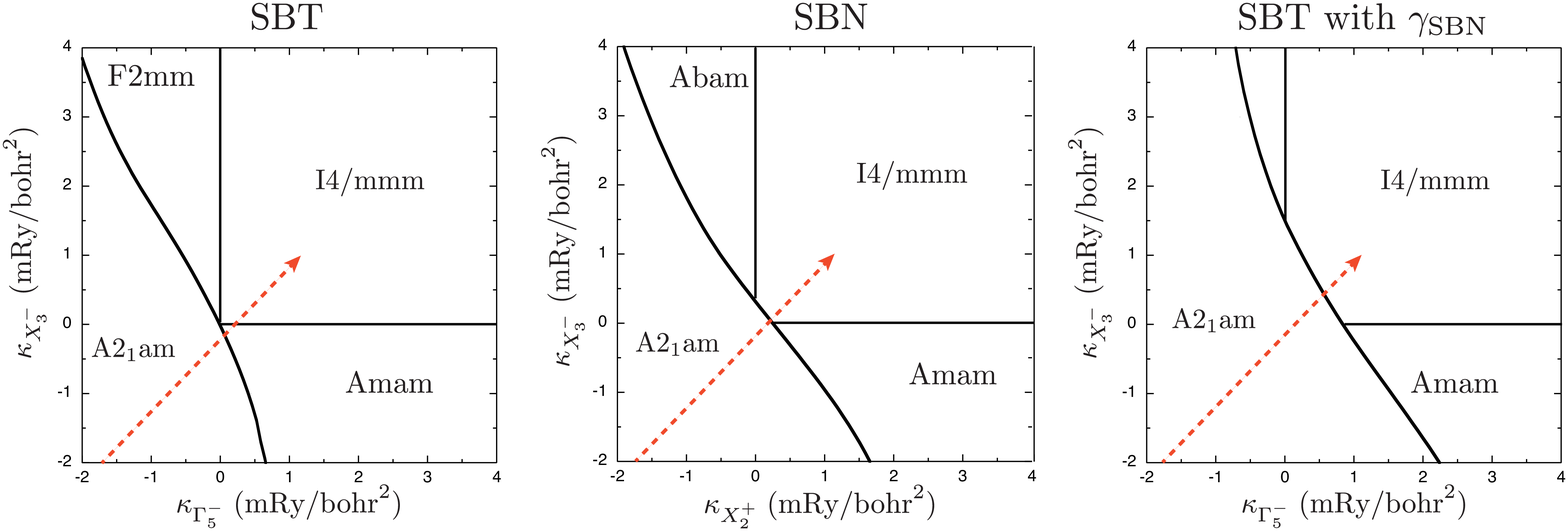}
\caption{Phase diagrams  for  SBT (left) and SBN (center). The variables of the abscises correspond to the second more unstable mode,  $\Gamma_5^-$  for SBT and $X_2^+$ for SBN. The diagram for  SBN presents a narrow gate that allows a direct transition from the ground state to the parent phase. The right panel corresponds to SBT but with the value of the trilinear coupling switched to that of SBN and it presents a wide passage that opens the possibility of a direct transition. The arrows show the same hypothetical paths along the three phase diagrams.}
\label{fig:fasediag_near}
\end{figure*}

Fig. \ref{fig:fasediag_near}(a) and (b)  show the phase diagrams of the two compounds in the space of the two more negative stiffness constants, $\kappa_{X_3^-}$ and $\kappa_{\Gamma_5^-}$ for SBT, and $\kappa_{X_3^-}$ and $\kappa_{X_2^+}$ for SBN. The diagrams correspond to a particular section where the difference between the stiffness constants $\kappa_{\Gamma_5^-}$ and $\kappa_{X_2^+}$ remains constant, which implies a similar temperature renormalization in both order parameters. The stability regions of Fig. \ref{fig:fasediag_near} correspond to the parent phase of symmetry $I4/mmm$ (no modes are frozen), the $A2_1am$ ground state (the three modes are frozen), and three phases associated with the condensation of a single mode:  $Amam$ for $X_3^-$,  $F2mmm$ for $\Gamma_5^-$ and $Abam$ for $X_1^+$. 

The phase diagram of SBN presents a finite gate that joins the stability regions of the tetragonal and ferroelectric phases. Thus, an appropriate renormalization of the order parameters  could drive the system directly from the ferroelectric state to the parent structure without an intermediate phase. However,  the topology of the phase diagram for SBT forbids this possibility and the intermediate phase must be present in a finite range of temperatures. This result could explain the existence of an \emph{avalanche} phase transition only in the case of SBN.

The fundamental influence of the strength of the trilinear coupling in the features of the phase diagrams is shown in Fig  \ref{fig:fasediag_near}(c). It corresponds to SBT but with the trilinear coupling constant changed to that of SBN,  that is, $\gamma=-22.55\;$mRy/bohr$^3$ instead of $\gamma=-13.81\;$mRy/bohr$^3$. The higher value of the coupling  induces a wide direct passage that allows an \emph{avalanche} phase transition. This suggests that the trilinear coupling does not only stabilize the ground states of the two compounds, but it also governs their sequence of phase transitions. 

\section{\label{sec:sei} Dimensionality of the order parameters and trilinear coupling}
The expressions of the previous section are not completely general since they do not take into account the correct dimensionality of the the order parameters. As usual in such calculations, the expansion of the energy in terms of distortions from  the high-symmetry configuration has been done by fixing the orientation of the multidimensional order parameters  along a specific direction, the direction that corresponds to one of the observed domains of the low-temperature structure.  In the case of SBT and SBN the three order parameters, $X_3^-$, $\Gamma_5^-$ and $X_2^+$, are two-dimensional and the experimental distortion corresponds to the specific direction indicated in Table \ref{tab:mode_decomposition}  and the symmetry-equivalent ones.

A general  distortions  must be expressed as two-dimensional vectors
and, in order to simplify the notation, we will use the following correspondences in cartesian and polar coordinates:
\begin{eqnarray*}
\mathbf{Q}_{X_3^-} & = &\bm{\varphi} \equiv (\varphi_1,\varphi_2)	=
 (\rho_\varphi \cos \theta_\varphi, \rho_\varphi \sin \theta_\varphi)\\
\mathbf{Q}_{\Gamma_5^-} & = &\bm{\psi} \equiv (\psi_1,\psi_2)	= 
(\rho_\psi \cos \theta_\psi, \rho_\psi \sin \theta_\psi)\\
\mathbf{Q}_{X_2^+} & = & \bm{\eta} \equiv (\eta_1,\eta_2) 	= (\rho_\eta \cos \theta_\eta, \rho_\eta \sin \theta_\eta)
\end{eqnarray*}

The invariant polynomials for general directions of the order parameters  obtained with the aid of the INVARIANTS
\cite{Invariants} utility are given in Table \ref{tab:invariants} in cartesian and polar coordinates. The lowest order term that includes anisotropy is the third order invariant related to the trilinear coupling;  moreover, the presence of this term is enough to explain the observed ground state. For a negative trilinear coupling the absolute minimum corresponds to the experimentally observed directions of the order parameters (Table \ref{tab:mode_decomposition}), whereas a positive coefficient should give a different structure with the same space group but $X_3^-$, $\Gamma_5^-$ and $X_2^+$ oriented along  $(a,-a)$, $(a,a)$ and  $(-a,a)$  respectively.

Following the approach of Ref. \onlinecite{Etxeb}, a simple microscopic Hamiltonian that retains the main features of  the  SBT/SBN system can be developed from an extended version of the two-dimensional $\phi^4$ model. The hamiltonian for a single isolated mode can be written as:

$$H_\phi=E_\phi \sum_i [|\bm{\phi}(i)|^2-1]^2+\frac{C_\phi}{2}\sum_{\langle i,j \rangle} [\bm{\phi}(i)-\bm{\phi}(j)]^2$$
with $\phi=\varphi, \psi, \eta$. $E_\phi$ is the depth of the double wells at sites $i$, $C_\phi$ are the harmonic couplings between nearest neighbours, and the ratio between both determines the behaviour of the system from order-disorder ($E>>C$) to displacive ($E<<C$). 

The  trilinear coupling can be included as an  on-site interaction:
\begin{eqnarray*}
H_{\varphi,\psi,\eta}&=&\frac{\gamma}{\sqrt{2}} \sum_i [\varphi_1(i)\psi_1(i) \eta_1(i)+\varphi_2 (i)\psi_1(i) \eta_2(i)- \\
&&   \varphi_2 (i)\psi_2(i) \eta_1(i)-\varphi_1(i)\psi_2(i) \eta_2(i)]
 \end{eqnarray*}
 and its contribution  at the absolute minimum is $-|\gamma| \rho_\varphi \rho_\psi \rho_\eta$ as in the one-dimensional case. In order to focus on the effect of the trilinear coupling term and to maintain the analogy with the one-dimensional case of Ref. \onlinecite{Etxeb}, we have not included in the Hamiltonian the rest of the fourth order terms that are allowed by symmetry. The complete Hamiltonian reads:
$$H=H_\varphi+H_\psi+H_\eta+H_{\varphi,\psi,\eta}$$

\begin{table}
\begin{ruledtabular}
\begin{tabular}{ll}
 Cart. coor.& \\
 \hline
2nd order & $|\bm{\varphi}|^2$, $|\bm{\psi}|^2$, $|\bm{\eta}|^2$  \\ \hline
3rd order & $\varphi_1\psi_1 \eta_1+\varphi_2 \psi_1 \eta_2- 
 \varphi_2 \psi_2 \eta_1-\varphi_1\psi_2 \eta_2$  \\ \hline
4th order &  $|\bm{\varphi}|^4$, $| \bm{\psi}|^4$, $|\bm{\eta}|^4$,  \\ 
&$\varphi_1^4+\varphi_2^4$, $\psi_1^4+\psi_2^4$, $\eta_1^4+\eta_2^4$, \\
 &  $| \bm{\varphi}|^2 | \bm{\psi} |^2$, $| \bm{\varphi} |^2 | \bm{\eta}|^2$, $| \bm{\psi} |^2 | \bm{\eta} |^2$, \\
  & $\varphi_1\varphi_2 \psi_1\psi_2$, $\varphi_1\varphi_2 \eta_1 \eta_2$ ,
 $\psi_1\psi_2 \eta_1\eta_2$  \\  \hline \hline
 Polar coor.& \\
 \hline
2nd order & $\rho_\varphi^2$, $\rho_\psi^2$, $\rho_\eta^2$  \\ \hline
3rd order & $\rho_\varphi \rho_\psi \rho_\eta [\cos \theta_\varphi \cos(\theta_\psi + \theta_\eta ) - 
\sin \theta_\varphi \sin(\theta_\psi - \theta_\eta )]$  \\ \hline
4th order & $\rho_\varphi^4$, $\rho_\psi^4$, $\rho_\eta^4$,  \\ 
& $\rho_\varphi^4(3+\cos 4\theta_\varphi)$, 
$\rho_\psi^4(3+\cos 4\theta_\psi)$, $\rho_\eta^4(3+\cos 4\theta_\eta)$, \\
 & $\rho_\varphi^2\rho_\psi^2$, $\rho_\varphi^2\rho_\eta^2$, $\rho_\psi^2\rho_\eta^2$, \\
& $\rho_\varphi^2 \rho_\psi^2\sin 2 \theta_\varphi \sin 2 \theta_\psi$, 
$\rho_\varphi^2 \rho_\eta^2\sin 2 \theta_\varphi \sin 2 \theta_\eta$, \\
&$\rho_\psi^2 \rho_\eta^2 \sin 2 \theta_\psi \sin 2 \theta_\eta$ \\
\end{tabular}
\caption{Invariant polynomials up to fourth order of the components of the order parameters in cartesian and polar coordinates. $\varphi$, $\psi$ and $\eta$ correspond to $X_3^-$, $\Gamma_5^-$ and $X_2^+$ respectively.}
\label{tab:invariants}
\end{ruledtabular}
\end{table}

The  analysis of all the possible combinations of the parameters of the Hamiltonian is beyond the scope of this work and  we have limited the study to some particular cases. First, we have considered the same displacive/order-disorder degree for the three order parameters ($E_\varphi/C_\varphi=E_\psi/C_\psi=E_\eta/C_\eta=C/E$) and two cases: order-disorder ($E/C=10$) and displacive ($E/C=0.1$). The dominance of the instability of the $X_3^-(\equiv \varphi)$ distortion in both compounds has been taken into account by assuming $C_\psi=C_\eta=0.8 C_\varphi$, and in consequence  $E_\psi=E_\eta=0.8 E_\varphi$. This choice yields the  same statistical behaviour for $\psi$ and $\eta$ and the relation $T_\psi=T_\eta=0.8 T_\varphi$ for the transition temperatures of the pure modes, i.e., without the trilinear coupling.

\begin{figure}
\includegraphics[scale = 0.5]{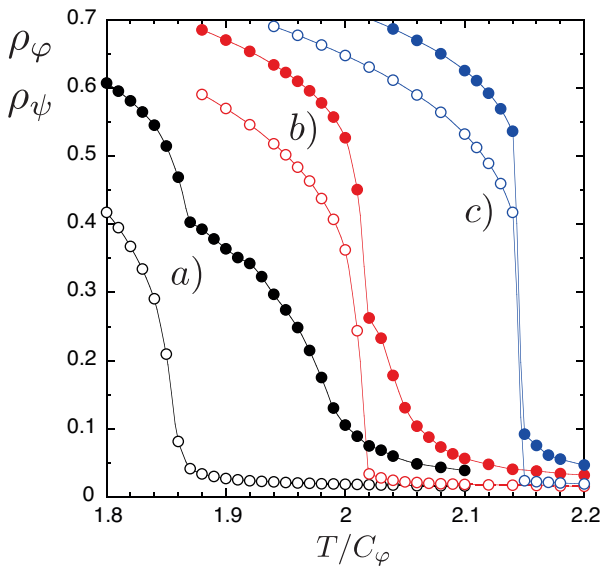} 
\caption{Amplitudes of the order parameters $\varphi$ (solid circles) and $\psi$ (open circles)  for $E/C=10$ and three values of the trilinear coupling:  a)
$\gamma = 0.1E_\varphi$,  b) $\gamma = 0.2E_\varphi$ and  c) $\gamma = 0.3E_\varphi$.  The amplitudes of $\eta$ are the same as  those of $\psi$ and are not shown in the figure.}
\label{fig:denakmc}
\end{figure}

We have studied the statistical properties of the model  by Monte Carlo simulations.
Calculations were carried out using cubic supercells containing $24\times 24\times 24$ sites and statistics were collected during at least $10^6$ Monte Carlo steps after a proper equilibration. Fig. \ref{fig:denakmc} shows the equilibrium amplitudes of the order parameters for the order-disorder case ($E/C=10$) and three different values of the trilinear coupling. Temperature is given  in units of $C_\varphi$. Three different regimes can be observed, for low values of $\gamma$ [Fig. \ref{fig:denakmc} (a)] both transitions are continuous and the range of stability of the intermediate phase is quite wide; for intermediate values of the trilinear coupling [Fig. \ref{fig:denakmc} (b)] the intermediate phase is stable in a narrow range of temperatures and the low temperature transition is first-order, and finally,  for stronger couplings [Fig. \ref{fig:denakmc} (c)] the intermediate phase disappears and a direct discontinuous \emph{avalanche} phase transition between the tetragonal and ferroelectric phase occurs with the simultaneous condensation of the three distortions.

\begin{figure*}
\includegraphics[scale = 0.45]{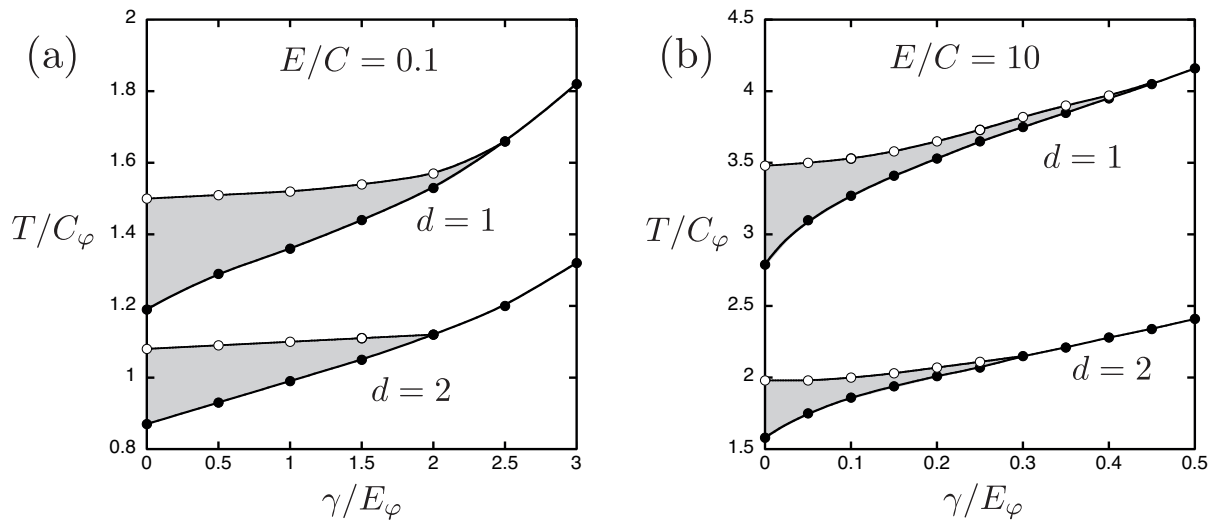} 
\caption{Transition temperatures of the one- and two-dimensional order parameters in terms of $\gamma$ for a) order-disorder and b) displacive systems obtained by Monte Carlo simulations. The grey area shows the stability region of the intermediate phase.}
\label{fig:fasediagmc}
\end{figure*}

Although results are qualitatively similar to those of previous works with one-dimensional order parameters,  the different calculation methods, mean field approximation and Monte Carlo simulations, prevent a quantitative evaluation of the role of the dimensionality in the stability of the phases. Therefore, we have also performed Monte Carlo simulations of the one-dimensional model and the comparison between the phase diagrams for both cases  are shown in Fig. \ref{fig:fasediagmc}. An accurate estimation of the transition temperatures was obtained by inspecting the probability distributions of the amplitudes of the distortions with simulations close to the critical regions. In the displacive regime the intermediate phase is suppressed for $\gamma\sim 2.5 E_\varphi$ and $\gamma\sim 2.0 E_\varphi$ for one and two-dimensional order parameters respectively [Fig.  \ref{fig:fasediagmc} (a)]. The influence of the dimensionality is more remarkable in the order-disorder regime where the \emph{avalanche} phase transition appears for $\gamma\sim 0.45 E_\varphi$ in the 1d case, and $\gamma\sim 0.3 E_\varphi$ when the multidimensionality of the order parameters is considered. It can be concluded that for higher dimensionalities of the irreducible representations associated to the order parameters the  trilinear coupling is more efficient to favour a direct phase transition between the high and low-temperature phases.

Multidimensional order parameters are ubiquitous in structural phase transitions. For instance, in completely different compounds as the ferroelectric  Ca$_3$Mn$_2$O$_7$ , antiferroelectric PbZrO$_3$  (PZO) and the double perovskites Sr$_2M$WO$_6$ ($M=$Zn, Ca and Mg), that do not present any dielectric anomaly, the trilinear coupling is critical to stabilize the ground state as in SBT and SBN. Table \ref{tab:dimensions} shows the relevant irreducible representations and their dimensions. Whereas the case of Ca$_3$Mn$_2$O$_7$ is similar to the SBT/SBN compounds, the high dimensionality of the irreps in the case of PZO could favour the observed strongly-discontinuous transition observed upon cooling from the cubic phase.   

\begin{table}
\begin{ruledtabular}
\begin{tabular}{llc lc |c }
 Compound & &  &\\
 \hline
Ca$_3$Mn$_2$O$_7$ & $X_2^+$(2)& $X_3^-$(2) & $\Gamma_5^-$(2)\\ 
PbZrO$_3$ & $R_5^-$(3)& $\Sigma_2$(12) & $S_2$(12)\\ 
Sr$_2M$WO$_6$  & $\Gamma_4^+$(3)& $X_3^+$(3) & $X_5^+$(6)\\ 
SBT/SBN  & $X_3^-$(2) & $\Gamma_5^-$(2) & $X_2^+$(2)\\ 

\end{tabular}
\caption{Three examples of  the multidimensionality of  order parameters that couple trilinearly (the labels of the irreducible representations and their dimensions are listed). The $M$ cation in Sr$_2M$WO$_6$ corresponds to Zn, Ca and Mg}
\label{tab:dimensions}
\end{ruledtabular}
\end{table}

\section{\label{sec:sei} CONCLUSIONS}
Despite the different sequence of phase transitions  that SBT and SBN go through, first principles calculations shows that the nature and main features of the instabilities in both compounds are qualitatively very similar, and that most of the conclusions about SBT obtained in Ref. \onlinecite{Manu} can be extended to SBN. Among the three symmetry adapted distortions ($X_3^-$, $\Gamma_5^-$ and $X_2^+$) that take part to  drive the tetragonal $I4/mmm$ parent phase to the polar $A2_1am$ structure, $X_3^-$ is dominant with the highest amplitude and the deepest energy well ($\sim$ 11 mRy per formula unit). The depths of the secondary distortions $\Gamma_5^-$ and  $X_2^+$ are similar in SBN ($\sim$ 3.5 mRy p.f.u.), while in SBT $X_2^+$ is hardly unstable and $\Gamma_5^-$ slightly deeper (5.2 mRy p.f.u.).

In both compounds , the simultaneous condensation of the  primary instability $X_3^-$ with any of the other two distortions is penalized energetically by positive  strong biquadratic couplings. Thus, the trilinear coupling is the key ingredient that allows the simultaneous condensation of the three modes and its role is essential to stabilize the observed ground state. 

The most noticeable difference between SBT and SBN is  associated with the magnitude of the trilinear term. The  polynomial expansion of the energy of both compounds shows that the trilinear coupling is  much stronger in SBN than in SBT, suggesting that its magnitude could be crucial to suppress the intermediate phase in SBN. Moreover, the analysis of phenomenological  phase diagrams for SBT and SBN shows that a higher value of this constant is enough to change the topology of the phase diagram, allowing a direct phase transition from the parent phase to the ground state.

Finally,  a simplified $\phi^4$ Hamiltonian that retains the symmetry requirements  and the correct dimensionality of the order parameters  has been developed for the SBT/SBN system. Monte Carlo simulations do not show qualitatively differences in comparison with the one-dimensional case\cite{Etxeb}:  the different sequence of phase transitions observed in both compounds can be reproduced by changing the strength of the trilinear coupling. However, according to the present calculations, the increment of the fluctuations associated with the higher dimensionality of the order parameters tends to favour the suppression of the intermediate phase and to reinforce the first order character of the direct transition between the high- and low-temperature structures. 
 

\begin{acknowledgments}
This work has been supported by  the Basque Government (Project No. IT-779-13) and the Spanish Ministry of Education and Science (Project  No. MAT2012-34740). One of the authors (U.P.) thankfully acknowledges the financial support of the University of the Basque Country (UPV/EHU).
\end{acknowledgments}

\end{document}